\documentclass[a4paper,UKenglish,cleveref, autoref, thm-restate]{lipics-v2021}

\title{The Central Valuations Monad}

\author{Xiaodong Jia}{School of Mathematics, Hunan University, Changsha, 410082, China}{}{}{}
\author{Michael Mislove}{Department of Computer Science, Tulane University, New Orleans, LA, USA}{}{}{}
\author{Vladimir Zamdzhiev}{Université de Lorraine, CNRS, Inria, LORIA, F 54000 Nancy, France}{}{}{}

\authorrunning{X. Jia, M. Mislove and V. Zamdzhiev} 

\Copyright{Xiaodong Jia, Michael Mislove  and Vladimir Zamdzhiev} 



\category{Early Ideas}

\relatedversion{} 

\nolinenumbers 

\EventEditors{Fabio Gadducci and Alexandra Silva}
\EventNoEds{2}
\EventLongTitle{9th Conference on Algebra and Coalgebra in Computer Science (CALCO 2021)}
\EventShortTitle{CALCO 2021}
\EventAcronym{CALCO}
\EventYear{2021}
\EventDate{August 31--September 3, 2021}
\EventLocation{Salzburg, Austria}
\EventLogo{}
\SeriesVolume{211}
\ArticleNo{18}

\newcommand{\rom}[1]{\rm{\uppercase\expandafter{\romannumeral #1}}}
\newcommand{\id}{\mathrm{id}}

\newcommand{\set}[2]{\{#1\mid#2\}}

\newcommand{\SSS}{{\mathcal S}}
\newcommand{\ZZ}{{\mathcal Z}}
\newcommand{\MM}{{\mathcal M}}
\newcommand{\WW}{{\mathcal W}}
\newcommand{\PP}{{\mathcal P}}
\newcommand{\VV}{{\mathcal V}}
\newcommand{\dcpo}{\mathbf{DCPO}}
\newcommand{\dom}{\mathbf{DOM}}
\newcommand{\defeq}{\stackrel{\textrm{{\scriptsize def}}}{=}}

\begin{document}

\maketitle

\begin{abstract}
We give a commutative valuations monad $\ZZ$ on the category $\dcpo$ of dcpo's and Scott-continuous functions. Compared to the commutative valuations monads given in \cite{monad-m}, our new monad $\ZZ$ is larger and it contains all push-forward images of valuations on the unit interval $[0, 1]$ along lower semi-continuous maps. We believe that this new monad will be useful in giving domain-theoretic denotational semantics for statistical programming languages with continuous probabilistic choice. 
\end{abstract}

\section{Introduction}
The valuations monad $\VV$ on the category $\dcpo$ of dcpo's and Scott-continuous functions is a staple of the domain-theoretic approach for denotational semantics of programming languages with probabilistic choice and recursion \cite{JonesP89,jones90}.
For a dcpo $D$, $\VV D$ consists of \emph{subprobability valuations} on $D$, which are the Scott-continuous functions $\nu$ from the set $\sigma D$ of Scott open subsets of $D$ to $[0, 1]$ satisfying \emph{strictness} ($\nu(\emptyset) = 0$) and \emph{modularity} ($\nu(U) + \nu (V) = \nu (U\cup V) +\nu(U\cap V)$). The set $\VV D$ is a dcpo in  the \emph{stochastic order}: $\nu_{1} \leq  \nu_{2}$ if and only if $\nu_{1}(U) \leq \nu_{2}(U)$ for all $U\in \sigma D$.
The \emph{unit} of $\VV$ at dcpo $D$ is the map $\eta_{D}\colon D \to \VV D :: x \mapsto \delta_{x}$, where $\delta_x$ is the \emph{Dirac valuation} at $x$, defined by $\delta_{x}(U) = 1$ if $x\in U$ and $\delta_{x}(U) = 0$ otherwise. For a Scott-continuous map $f\colon D\to \VV E$, the \emph{Kleisli extension} $f^{\dagger}$ of $f$ is defined by $f^{\dagger}(\nu)(U) = \int_{x\in X} f(x)(U) d \nu$ for $\nu\in \VV D$ and $U\in \sigma E$. The integral in this definition is a Choquet type integral: for a general Scott-continuous function $h\colon D \to [0, 1]$, the value of $\int_{x\in X} h d\nu$ is defined to be the Riemann integral $\int_{0}^{1} \nu(h^{-1}(t, 1]) d t$. Following this, the action of $\VV$ on a Scott-continuous function $g\colon D\to E$ between dcpo's $D$ and $E$ is $\VV(g) \defeq (\eta_{E}\circ g)^{\dagger}$; concretely, 
for $\nu\in \VV D$ and $U\in \sigma E$, $\VV(g)(\nu)(U) = \nu(g^{-1}(U))$. Subprobability valuations on general topological spaces and the corresponding integral of lower semi-continuous functions against subprobability valuations can be defined similarly~\cite{JonesP89}. 

While it is well-known that $\VV$ can be restricted to a commutative monad on the category $\dom$ of domains and Scott-continuous functions, it is unknown whether $\VV$ can be restricted to any Cartesian closed full subcategory of $\dom$. This is known as the \emph{Jung-Tix problem}~\cite{jungtix}.

One may note that the category $\dcpo$ itself is Cartesian closed and $\VV$ is a monad on it. What does one lose if we use the category $\dcpo$ and monad $\VV$ for semantics? A short answer is that compared to $\dom$, $\VV$ is not known to be \emph{commutative} over $\dcpo$, which is an important property for the denotational semantics of programming languages.
Commutativity of $\VV$ over $\dcpo$ is equivalent to showing the following Fubini-style equation
\begin{equation}\label{fubini}
 \int_{x\in D} \int_{y\in E} h(x, y) d\mu d\nu = \int_{y\in E} \int_{x\in D} h(x, y) d\nu d\mu
\end{equation} 
holds for all dcpo's $D$ and $E$, all Scott-continuous functions $h\colon D\times E \to [0, 1]$ and all $\nu \in \VV D, \mu\in \VV E$. As pointed out in~\cite{monad-m}, the main difficulty in establishing \eqref{fubini} over $\dcpo$ is that the Scott topology on the product dcpo $D\times E$ may be different from the product topology $\sigma D\times \sigma E$. Actually,
we do know that Equation~(\ref{fubini}) holds for those functions $h$ that are continuous when $D\times E$ is given the product topology $\sigma D\times \sigma E$ (\cite[Lemma 2.37]{jones90}).

Instead of directly proving \eqref{fubini}, we showed (together with Lindenhovius) how to construct three submonads of $\VV$ that are commutative on $\dcpo$, and used each one to give a sound and (strongly) adequate semantics to $\mathtt{PFPC}$ (Probabilistic FixPoint Calculus) \cite{monad-m}.
The simplest of those three monads is the monad $\MM$. For each dcpo $D$, $\MM D$ is defined to be the smallest sub-dcpo of $\VV D$ that contains $\SSS D$, the family of \emph{simple valuations} on $D$, where a simple valuation is a finite convex sum of Dirac valuations.  The other two commutative monads are denoted $\WW$ and $\PP$ and the following inclusions hold for each dcpo $D: \SSS D \subseteq \MM D \subseteq \WW D \subseteq \PP D \subseteq \VV D$.

Each of our three monads is large enough to interpret \emph{discrete} probabilistic choice in $\mathtt{PFPC}$~\cite{monad-m}. However, it is unclear if any of these monads is large enough to interpret \emph{continuous} probabilistic choice.
In this note, we define a new commutative valuations monad $\ZZ$ on the category $\dcpo$ which is larger than $\MM, \WW$ and $\PP$ with the hope of addressing this problem.

\section{Central Valuations}

Our idea for defining $\ZZ$ is inspired by the notion of centre in group theory (which always forms an abelian subgroup) and the notion of centre of a premonoidal category (which always forms a monoidal subcategory) \cite{premonoidal}.

\begin{definition}
A subprobability valuation $\nu$ on a dcpo $D$ is called a \emph{central valuation} if for any dcpo $E$, any valuation $\mu$ on $E$, and any Scott-continuous function $h\colon D\times E \to [0, 1]$, we have 
$$ \int_{x\in D} \int_{y\in E} h(x, y) d\mu d\nu = \int_{y\in E} \int_{x\in D} h(x, y) d\nu d\mu. $$
  We shall write $\ZZ D$ for the set of all central valuations on a dcpo $D$.
\end{definition}

It is easy to see that simple valuations are central, and that the central valuations are closed under directed suprema under the stochastic order. Thus, for each dcpo $D$, $\ZZ D$
is a sub-dcpo of  $\VV D$ containing $\SSS D$. Moreover, we have the following theorem, which can be proved using the \emph{disintegration formula} in \cite{goubault21}.  

\begin{theorem}
  The assignment $\ZZ(-)$ extends to a commutative monad over the category $\dcpo$ when equipped with the (co)restricted monad operations of $\VV$. In other words, $\ZZ$ is a commutative submonad of $\VV$.
\end{theorem}
\begin{proof}
The unit of $\ZZ$ at dcpo $D$ sends each $x\in D$ to $\delta_{x}$ which is obviously a central valuation.

Let $f\colon C\to \ZZ D$ be a Scott-continuous function. Then $f$ can also be viewed as a Scott-continuous map from $C$ to $\VV D$, since $\ZZ D$ is a sub-dcpo of $\VV D$. We prove that $f^{\dagger} \colon \VV C \to \VV D$ maps central valuations on $C$ to central valuations on $D$. Towards this end, we pick $\mu$ from $\ZZ C$, and assume that $E$ is a dcpo, $\nu$ is an arbitrary subprobability valuation on $E$ and $h\colon D\times E \to [0, 1]$ is a Scott-continuous map.  Then by the disintegration formula (see Lemma~3.1(iii) in~\cite{goubault21}) we have that
$$ \int_{y\in E} \int_{x\in D} h(x, y) d(f^{\dagger}(\mu)) d\nu = \int_{y\in E} \int_{t\in C}\int_{x\in D} h(x, y) df(t)d\mu d\nu,$$
and the right side of the equation is equal to 
\[\int_{t\in C} \int_{x\in D}\int_{y\in E} h(x, y)d\nu df(t)d\mu \]
by the fact that $f(t), t\in D$ and $\mu$ are central valuations. 
Again, by the disintegration formula that is just $\int_{x\in D}\int_{y\in E} h(x, y)d\nu df^{\dagger}(\mu)$.
Hence we have proved that $f^{\dagger}(\mu)$ is indeed a central valuation provided that $\mu$ is. 
Similar arguments show that the monadic strength also (co)restricts as required.
The corresponding (co)restrictions of the monadic operations of $\VV$ to $\ZZ$ validate that $\ZZ$ is a strong monad on $\dcpo$. The commutativity of $\ZZ$, which is equivalent to Equation~\eqref{fubini} holding for all dcpo's $D$ and $E$ and central valuations $\mu$ and $\nu$ on them, is then obvious by definition of $\ZZ$.
\end{proof}

In fact, it is proved in \cite{monad-m} that all \emph{point-continuous valuations} are central and therefore
$ \SSS D \subseteq \MM D \subseteq \WW D \subseteq \PP D \subseteq \ZZ D \subseteq \VV D$ for each dcpo $D$. Therefore $\ZZ$ is the largest commutative submonad of $\VV$ known so far.
Furthermore, observe that $\ZZ = \VV$ iff $\VV$ is a commutative monad on $\dcpo.$ The latter has been an open problem since 1989, and our simple observation leads us to believe $\ZZ$ is a very large commutative submonad of $\VV$.

It is not difficult to see that, in order to model sampling against continuous probability distributions on the interval $[0, 1]$, the monad used for the semantics should at least contain the push-forward images of the Lebesgue valuation on $[0, 1]$ (equipped with the metric topology) along lower semi-continuous maps.
We can demonstrate even more is true of our new monad $\ZZ$ (see Theorem~4 below). For this, let us first
recall that a space $X$ is called \emph{core-compact} if the set $\mathcal OX$ of all open subsets of $X$ is a continuous lattice in the inclusion order. Equivalently, $X$ is core-compact if and only if for each open subset $U$ of $X$ and $x\in U$, there exists an open subset $V$ such that $x\in V\ll U$, where $V\ll U$ means that $V$ is \emph{way-below} $U$ in the sense of domain theory. 
Many important spaces are core-compact. For example, each locally compact space is core-compact, and in particular, the unit interval with the usual topology is compact Hausdorff, hence locally compact hence core-compact.

\begin{lemma}\label{lemma:core-compact-to-dcpo}
Let $X$ be a core-compact topological space. Let $D, E$ be dcpos, and $f\colon X\to D$ a lower semi-continuous map, i.e., $f$ is continuous when $D$ is equipped with the Scott topology. Then the map $f \times \id_{E} \colon X \times \Sigma E \to D \times E$ is also lower semi-continuous, where $\Sigma E$ denotes the topological space $(E, \sigma E)$ and 
  $X \times \Sigma E$ is the topological product of $X$ and $\Sigma E$. 
\end{lemma}
\begin{proof}
First, we assume that $X$ is core-compact and prove that $f\times \id_{E}$ is lower semi-continuous. Towards this end, we pick a Scott open subset $O$ of $D\times E$, and assume that $f \times \id_{Y}((x_{0}, e_{0})) \in O$, that is $(f( x_{0}), e_{0}) \in O$. We must find an open neighbourhood $U$ of $x_{0}$ in $X$ and a Scott open neighbourhood $V$ of $e_{0}$ in $E$ such that $f\times \id_{Y}(U\times V ) \subseteq O$. 
We let $A = \set{x\in X}{(f(x), e_{0}) \in O }$. Then $A$ is just $f^{-1}(O_{e_{0}})$, where $O_{e_{0}} =\set{d\in D}{ (d, e_{0}) \in O } $. 
Since $f\colon X\to D$ is lower semi-continuous and $O_{e_{0}}$ is Scott open in $D$, we know that $A$ is an open neighbourhood of $x_{0}$ in $X$. 
Now the core-compactness of $X$ enables us to find, in $X$, an open subset $U$ and a sequence of open subsets $U_{i}, i \in \mathbb N$ such that 
$x_{0}  \in U\ll \cdots \ll U_{n} \cdots \ll U_{1} \ll A $. For each $U_{n}, n\in \mathbb N$, we define $V_{n} = \set{e}{ f(x, e) \in O ~\text{for all $x\in U_{n}$}}$ and let $V = \bigcup_{n\in \mathbb N} V_{n}$. Since for each $n\in \mathbb N$, $U_{n} \subseteq A$, we have for all $x\in U_{n}$, $(f(x), e_{0})\in O$. Thus we know that $e_{0} \in V_{n}$ for each $n\in \mathbb N$, and hence $e_{0} \in V$. Moreover, for any $(x, e)\in  U\times  V$, there exists a natural number $n$ such that $e\in V_{n}$, then it follows that $f\times \id_{E} ((x, e)) = (f(x), e) \in f(U) \times V_{n} \subseteq f(U_{n}) \times V_{n} \subseteq O$. The last inclusion is due to the construction of $V_{n}$. To sum up, it is true that $f\times \id_{E} (U\times V) \subseteq O$. Since $U$ is an open subset of $X$ which contains $x_{0}$ and $e_{0} \in V$, we finish the proof by showing that $V$ is  Scott open in $E$. To this end we let
 $\{e_{i}\}_{i\in I}$ be a directed subset of $E$ with $\sup_{i\in I} e_{i} \in V$. For each $i\in I$, set $W_{i} = \set{ x\in X }{(f(x), e_{i}) \in O}$. It is easy to see that $\set{W_{i}}{ i\in I}$ is a directed family of open subsets of $X$. Since $\sup_{i\in I} e_{i} \in V = \bigcup_{n\in \mathbb N}V_{n}$, $\sup_{i\in I}e_{i}$ is in some $V_{n}$. This means that for each $x\in U_{n}$, $(f(x), \sup_{i\in I} e_{i}) \in O$. Because $O$ is Scott open,  for each $x\in U_{n}$, there exists $i \in I$ such that $(f(x), e_{i} ) \in O$, i.e., $x\in W_{i}$. Hence we have that $\sup_{i\in I} e_{i} \in V_{n} \subseteq \bigcup_{i\in I}W_{i}$. Remember that $U_{n+1} \ll U_{n}$, it follows that $U_{n+1} \subseteq W_{j}$ for some $j\in I$. By definition of $W_{j}$, we know that $f(U_{n+1}) \times  \{ e_{j }\} \subseteq O$, which means that $e_{j} \in V_{n+1}$, this time by definition of $V_{n+1}$. So we find $j\in I$ with $e_{j} \in V_{n+1} \subseteq V$, and indeed $V$ is Scott open in $E$. 
\end{proof}

\begin{theorem}\label{mainth}
Let $X$ be a core-compact space and $f$ be a lower semi-continuous map from $X$ to a dcpo $D$. If $\nu$ is a valuation on $X$, then 
$f_{*}(\nu) \defeq \lambda O\in \sigma D. \nu(f^{-1}(O))$, the \emph{push-forward valuation along $f$}, is a central valuation on $D$. In particular, for a core-compact dcpo $D$, all valuations on $D$ are central, i.e., $\VV D = \ZZ D$. 
\end{theorem}

\begin{proof}
By definition, we prove for any dcpo $E$, continuous valuations $\mu$ on $E$ and Scott-continuous map $h\colon D\times E \to [0, 1]$ the equation 
$$ \int_{x\in D} \int_{y\in E} h(x, y) d\mu d f_{*}(\nu) = \int_{y\in E} \int_{x\in D} h(x, y) d f_{*}(\nu) d\mu$$
holds. 

Note that for each $y\in E$, the map $g \defeq (x \mapsto \int_{y\in E} h(x, y) d\mu) \colon D \to [0, 1] $ is Scott-continuous, and $f\colon X \to D$ is lower semi-continuous. Hence for the left side of the above equation we have

$$\int_{x\in D} \int_{y\in E} h(x, y) d\mu d f_{*}(\nu) = \int_{x\in X} g(f (x))d\nu  =\int_{x\in X} \int_{y\in E} h(f(x), y) d\mu d\nu. $$

The first equality follows form the so-called \emph{change-of-variable} formula, which can be found in~\cite{jones90}.
As a consequence of it, we also have that  $$\int_{y\in E} \int_{x\in D} h(x, y) d f_{*}(\nu) d\mu  =  \int_{y\in E} \int_{x\in X} h(f(x), y) d\nu d\mu.$$ 
Since $X$ is core-compact and the function $f\colon X\to D$ is lower semi-continuous, by Lemma~\ref{lemma:core-compact-to-dcpo} we know that 
$f\times \id_{E} \colon X\times \Sigma E \to X\times Y$ is lower semi-continuous. 
This implies that the map $(x, y) \mapsto h(f(x), y) \colon X\times \Sigma E \to [0, 1]$ is lower semi-continuous. Hence by Lemma~2.37 in \cite{jones90} we know that $\int_{x\in X} \int_{y\in E} h(f(x), y) d\mu d\nu =  \int_{y\in E} \int_{x\in X} h(f(x), y) d\nu d\mu$, which finishes the proof. 

The second claim is a straightforward consequence of the first one. 
\end{proof}

\begin{theorem}
  Let $f : [0,1] \to D$ be a lower semi-continuous map into a dcpo $D$. If $\nu$ is any continuous valuation on $[0,1]$, then $f_{*}(\nu)$ is a central valuation on $D$.
\end{theorem}
\begin{proof}
Since $[0, 1]$ is core-compact in the usual topology, the result follows from Theorem~\ref{mainth}.
\end{proof}

We have not been able to establish the above theorem for any of the monads $\MM, \WW$ or $\PP$, so we believe that $\ZZ$ is a promising candidate for modeling \emph{continuous} probabilistic choice. We plan to address this in future work.

\section*{Acknowledgement}

We thank the anonymous reviewers for their feedback which led to improvements of this paper.
Xiaodong Jia acknowledges the support of NSFC (No. 12001181).

\bibliography{refs}

\begin{thebibliography}{1}

\bibitem{goubault21}
Jean Goubault-Larrecq, Xiaodong Jia, and Cl\'ement Th\'eron.
\newblock {A Domain-Theoretic Approach to Statistical Programming Languages},
  2021.
\newblock Preprint.
\newblock URL: \url{https://arxiv.org/abs/2106.16190}.

\bibitem{monad-m}
Xiaodong Jia, Bert Lindenhovius, Michael Mislove, and Vladimir Zamdzhiev.
\newblock Commutative monads for probabilistic programming languages.
\newblock In {\em Logic in Computer Science ({LICS 2021})}, 2021.
\newblock \href {http://arxiv.org/abs/2102.00510} {\path{arXiv:2102.00510}}.

\bibitem{JonesP89}
C.~Jones and Gordon~D. Plotkin.
\newblock A probabilistic powerdomain of evaluations.
\newblock In {\em Proceedings of the Fourth Annual Symposium on Logic in
  Computer Science {(LICS} '89), Pacific Grove, California, USA, June 5-8,
  1989}, pages 186--195. {IEEE} Computer Society, 1989.
\newblock \href {https://doi.org/10.1109/LICS.1989.39173}
  {\path{doi:10.1109/LICS.1989.39173}}.

\bibitem{jones90}
Claire Jones.
\newblock {\em {Probabilistic Non-determinism}}.
\newblock PhD thesis, University of Edinburgh, {UK}, 1990.
\newblock URL: \url{http://hdl.handle.net/1842/413}.

\bibitem{jungtix}
Achim Jung and Regina Tix.
\newblock The troublesome probabilistic power domain.
\newblock In {\em Comprox III, Third Workshop on Computation and
  Approximation}, volume~13, pages 70 -- 91, 1998.

\bibitem{premonoidal}
John Power and Edmund Robinson.
\newblock {Premonoidal Categories and Notions of Computation}.
\newblock {\em Math. Struct. Comput. Sci.}, 7(5):453--468, 1997.
\newblock \href {https://doi.org/10.1017/S0960129597002375}
  {\path{doi:10.1017/S0960129597002375}}.

\end{thebibliography}

\end{document}